\documentclass[fleqn,conference]{IEEEtran}
\IEEEoverridecommandlockouts
\usepackage[utf8]{inputenc}
\usepackage{cite}
\usepackage{url}
\usepackage{amsmath,amssymb,amsfonts}
\usepackage[ruled,vlined,linesnumbered]{algorithm2e}
\DeclareMathOperator*{\argmin}{arg\,min}
\usepackage{algorithmic}
\usepackage{graphicx}
\usepackage{textcomp}
\usepackage{tikz}
\usetikzlibrary{arrows.meta, positioning, fit, backgrounds}
\usepackage[x11names,dvipsnames,table]{xcolor}
\usepackage{comment}
\def\BibTeX{{\rm B\kern-.05em{\sc i\kern-.025em b}\kern-.08em
    T\kern-.1667em\lower.7ex\hbox{E}\kern-.125emX}}
\usepackage{multirow}
\usepackage{tikz}
\usetikzlibrary{decorations.pathreplacing}
\usetikzlibrary{matrix}
\usepackage{pgfplots}
\pgfplotsset{compat=1.18}
\pgfplotsset{every axis legend/.append style={font=\tiny}}
\usepackage{pgfplots}
\pgfplotsset{compat=1.18}
\usepackage{xcolor}
\usepackage{tikz}
\usetikzlibrary{arrows.meta, positioning, fit, backgrounds, calc}

\definecolor{navyblue}{RGB}{31, 56, 100}
\definecolor{orange}{RGB}{255, 165, 0}
\usepackage{pgfplots}
\usepgfplotslibrary{statistics}
\pgfplotsset{compat=1.18}

\usepackage{booktabs}           
\usepackage{tabularx}           
\usepackage{multirow}           
\usepackage{array}              

\usepackage{graphicx}
\usepackage{tikz}
\usepackage{caption}
\usepackage{subcaption}         

\graphicspath{{images/runs/run_20260522_220151/}}

\usepackage[font={small}]{caption, subfig}
\DeclareMathSizes{10}{9}{6}{5}

\usepackage[nonumberlist]{glossaries}
\usepackage{mdframed} 
\usepackage{multicol} 
\usepackage{enumitem} 

\definecolor{rangeA1}{RGB}{255,255,179} 
\definecolor{rangeA2}{RGB}{255,255,128} 
\definecolor{rangeA3}{RGB}{255,255,77}  
\definecolor{rangeA4}{RGB}{255,255,0}   
\definecolor{rangeA5}{RGB}{230,230,0}   
\definecolor{rangeA6}{RGB}{204,204,0}   

\definecolor{rangeB1}{RGB}{255,230,179} 
\definecolor{rangeB2}{RGB}{255,204,128} 
\definecolor{rangeB3}{RGB}{255,179,77}  
\definecolor{rangeB4}{RGB}{255,153,0}   
\definecolor{rangeB5}{RGB}{230,138,0}   
\definecolor{rangeB6}{RGB}{204,122,0}   

\definecolor{rangeC1}{RGB}{255,179,179} 
\definecolor{rangeC2}{RGB}{255,128,128} 
\definecolor{rangeC3}{RGB}{255,77,77}   
\definecolor{rangeC4}{RGB}{255,0,0}     
\definecolor{rangeC5}{RGB}{204,0,0}     
\definecolor{rangeC6}{RGB}{153,0,0}     

\definecolor{rangeD1}{RGB}{179,255,179} 
\definecolor{rangeD2}{RGB}{128,255,128} 
\definecolor{rangeD3}{RGB}{77,255,77}   
\definecolor{rangeD4}{RGB}{0,255,0}     
\definecolor{rangeD5}{RGB}{0,204,0}     
\definecolor{rangeD6}{RGB}{0,153,0}     

\definecolor{rangeE1}{RGB}{179,179,255} 
\definecolor{rangeE2}{RGB}{128,128,255} 
\definecolor{rangeE3}{RGB}{77,77,255}   
\definecolor{rangeE4}{RGB}{0,0,255}     
\definecolor{rangeE5}{RGB}{0,0,204}     
\definecolor{rangeE6}{RGB}{0,0,153}     

\definecolor{exact}{RGB}{0,0,0} 

\newif\ifcomm
\commtrue 
\ifcomm
\else
\commfalse
\fi
\ifcomm
\newcommand\ak[1]{\textcolor{orange}{AK: #1}}
\newcommand\ogi[2]{\textcolor{magenta}{Ogi: #1}}

\else
\newcommand\ak[1]{}
\newcommand\ogi[2]{}
\fi

\newglossary{printedglossary}{not}{ntn}{Nomenclature}
\newglossary{internal}{sym}{sbl}{internal wont be printed}

\makeglossaries

\newglossarystyle{customlist}{%
  \setglossarystyle{list}%
}

\setglossarystyle{customlist}
\begin{document}
\title{ST-NDT: A Topological Framework for Reducing Communication Overhead in Network Digital Twins}

\author{\IEEEauthorblockN{John Sengendo}
\IEEEauthorblockA{Dept. of Information Engineering and Computer Science\\
University of Trento\\
Trento, ITALY and C.N.I.T\\
Email: john.sengendo@unitn.it}
\and
\IEEEauthorblockN{Fabrizio Granelli}
\IEEEauthorblockA{Dept. of Information Engineering and Computer Science\\
University of Trento\\
Trento, ITALY and C.N.I.T \\
Email: fabrizio.granelli@unitn.it}

}

\maketitle

\begin{abstract}
\noindent Next-generation networks (NGN) are becoming increasingly complex, especially with the increasing size of topologies. To manage this complexity requires the adoption of frameworks that enable real-time monitoring, optimization, and ``what-if'' case scenario analysis. Network Digital Twins (NDTs) have emerged as a key enabler technology for supporting these capabilities because of their ability to provide digital replicas of physical networks operations, enabling scenario testing without interfering with the live network. Although NDTs are considered a key enabler technology, maintaining continuous synchronization with the Physical Twin (PT) introduces considerable communication overhead and excessive bandwidth utilization, limiting sustainability in resource-constrained networks. This paper proposes a topological signal processing framework for sparse network monitoring in NDTs. We represent network edge flows as signals defined on a cell complex, and exploit the Hodge spectral structure of the graph to identify a minimal set of maximally informative sensor edges, thereby reducing the measurement overhead between the physical network and its digital twin replica.
Results show that the proposed Sparse Topological Network Digital Twin (ST-NDT) framework consistently outperforms all four baseline frameworks across most of the edge monitoring budgets. For example, at a 20\% monitoring budget (48 of 243 edges monitored), ST-NDT achieves improvements of 10.85\%, 11.24\%, 7.36\%, and 6.15\% over degree-based, core-number, betweenness centrality, and random-selection frameworks, respectively.
\end{abstract}

\section{Introduction}

The emergence of data-intensive applications such as virtual reality, telemedicine, and autonomous vehicles places stringent requirements on modern networks such as 5G, particularly for low latency, high bandwidth, and high reliability \cite{orlosky2017virtual,dananjayan20215g}. As recent studies additionally underscore, the complexity of such environments particularly vehicular networks stems not only from application demands but also from the highly dynamic nature of mobility, which makes static resource allocation strategies insufficient \cite{sayegh2026toward}. Network Digital Twins (NDTs) have recently emerged as a powerful paradigm to enable networks operate smoothly under these requirements. As pointed out in literature, NDTs provide virtual replicas for physical networks \cite{Sanz}. During operation, they ingest telemetry measurements and use underling algorithms like machine learning (ML) models to emulate operations of the live network~\cite{almasan2022network}. By decoupling network operations and troubleshooting from the digital replica, NDTs enable proactive management, scenario analysis and accelerated deployment of new network services~\cite{almasan2022network}. However, for their efficient operation, synchronization is key to keep both the PT and DT aligned. Maintaining synchronization between the physical network and its digital counterpart is non-trivial, especially when it comes to monitored information needed to replicate back the topology and the complexity of the topologies in 5G and 6G networks \cite{9090214}. 

In vehicular digital twin networks for instance, to maintain synchronization, data must be exchanged both within each digital twin and among different twins. As topologies expand, so does the needed monitored information, thus creating a communication overhead bottleneck~\cite{VDT_Priority}. The challenge becomes even more critical in cloud–edge digital twin architectures, where real-time synchronization and analytics depend on frequent bidirectional communication, from data acquisition to control feedback between edge devices, and centralized cloud platforms. Prior works in \cite{mishra2023towards,kalasapura2023twinsync, 11352012} additionally noted that bandwidth (BW) limits, latency, jitter, and scalability constraints can all degrade synchronization accuracy, responsiveness, and reliability.

\begin{figure}[h]
    \centering
    \includegraphics[width=\columnwidth]{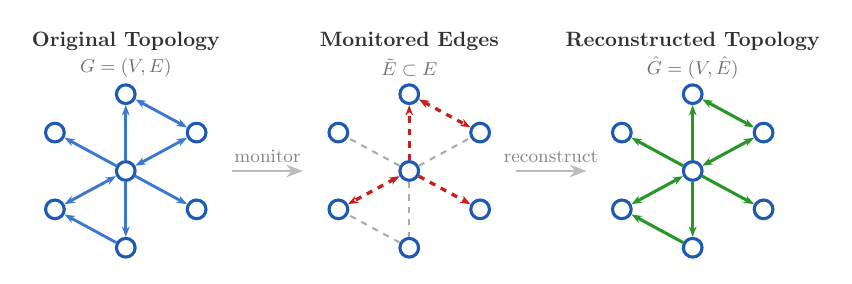}
    \caption{Network topology monitoring and reconstruction. The original topology $G=(V,E)$ is monitored over a subset of edges $\tilde{E} \subset E$.}
    \label{fig:topology}
\end{figure}

A promising approach to mitigating the overhead challenge while remaining within the BW requirements is to reduce the amount of information needed to maintain synchronization, while preserving the information content relevant for management tasks. Graph signal processing (GSP) provides frameworks for analyzing data defined on graphs, it generalizes classical signal processing concepts such as Fourier transforms, filtering and sampling to irregular domains~\cite{GSP_Overview}. 

Nevertheless, many network operations involve interactions among more than two nodes, multi-user interference patterns, and service slices, so pairwise graph models may be insufficient. Topological signal processing (TSP) further extends GSP to higher-order domains by considering signals on simplices (nodes, edges, triangles)~\cite{TSP_Simplicial}. It provides sampling, filtering and inference tools tailored to simplicial complexes~\cite{TSP_Simplicial}. Figure~\ref{fig:topology} illustrates the concept, demonstrating how a subset of monitored edges $\tilde{E} \subset E$ is sufficient to recover the full network topology $\hat{G}=(V,\hat{E})$ from the 
original graph $G=(V,E)$. Recent work further generalizes TSP to cellular complexes, minimizing the inclusion property required by simplicial complexes and yielding a richer algebraic structure with a better complexity/accuracy trade-off~\cite{TSP_Cell}. These advances suggest that network states can be modeled as higher-order topological signals, compressible via harmonic decompositions and sparse sampling which can be extendable in NDTs for replicating back network topologies.
This forms the focal of the key contributions our paper presents, which are listed as follows:
\begin{itemize}
  \item For realism, we leverage a real-world ISP backbone topology with 197 nodes and 243 edges.
  
  \item We exploit the Hodge spectral decomposition to design a topology-aware sensor placement strategy that spans all three Hodge subspaces under any monitoring budget.

  \item We benchmark the proposed ST-NDT framework against four baselines to provide a comparative performance analysis.

  \item We additionally underscore how ST-NDT is consistent with the 6G and 3GPP requirements.
\end{itemize}

The rest of this paper is organized as follows. Section~\ref{sec:related} reviews related work on network digital twins and topological signal processing. Section~\ref{sec:methodology} details the methodology of our work and section~\ref{sec:experimental_setup} presents the experimental setup. Section~\ref{sec:results} discusses the results. Finally, conclusions are drawn in section~\ref{sec:conclusion}.

\section{Related work}\label{sec:related}
In this section, we explore the current state-of-the-art inline with the proposed framework.
\subsection{Network Digital Twins, communication overhead and synchronization}

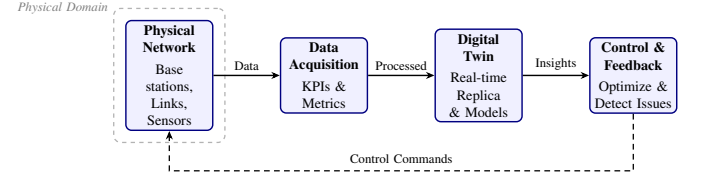
\begin{figure}[h]
    \centering
    \resizebox{\columnwidth}{!}{
%
\begin{tikzpicture}[
    font=\normalsize,
    box/.style={
        draw,
        rounded corners=4pt,
        minimum width=2.0cm,
        minimum height=1.5cm,
        text width=1.9cm,
        align=center,
        line width=1pt,
        fill=blue!6,
        draw=blue!50!black
    },
    solidarrow/.style={-{Stealth[length=6pt, width=5pt]}, line width=1pt},
    dasharrow/.style={-{Stealth[length=6pt, width=5pt]}, dashed, line width=0.9pt,
                      dash pattern=on 4pt off 3pt}
]
\node[box] (phys) at (0,   0) {%
    \textbf{Physical\\Network}\\[3pt]
    Base stations,\\Links, Sensors
};
\node[box] (data) at (3.8, 0) {%
    \textbf{Data\\Acquisition}\\[3pt]
    KPIs \& Metrics
};
\node[box] (twin) at (7.6, 0) {%
    \textbf{Digital\\Twin}\\[3pt]
    Real-time\\Replica \& Models
};
\node[box] (ctrl) at (11.4,0) {%
    \textbf{Control \&\\Feedback}\\[3pt]
    Optimize \&\\Detect Issues
};
\draw[solidarrow] (phys.east) --
    node[above, font=\small]{Data} (data.west);
\draw[solidarrow] (data.east) --
    node[above, font=\small]{Processed} (twin.west);
\draw[solidarrow] (twin.east) --
    node[above, font=\small]{Insights} (ctrl.west);
\draw[dasharrow]
    (ctrl.south)
    -- ++(0, -1.4)
    -- ++(-11.4, 0)
    -- (phys.south);
\node[font=\small, below] at (5.7, -1.8) {Control Commands};
\begin{pgfonlayer}{background}
    \node[
        draw=gray!60,
        dashed,
        rounded corners=5pt,
        inner sep=8pt,
        line width=0.9pt,
        label={[font=\small\itshape, gray]above left:Physical Domain}
    ] [fit=(phys)] {};
\end{pgfonlayer}
\end{tikzpicture}}
     \caption{NDT--PT interaction showing key steps. Data acquisition, digital twin replication, and closed-loop control.}
    \label{fig:ndt_pt}
\end{figure}

Digital twin technology has gained traction across engineering domains for modeling and replicating complex systems~\cite{almasan2022network}. In networking, the concept of a NDT aims to build accurate, real-time, data‑driven replicas of communication infrastructures to facilitate tasks such as performance evaluation, routing optimization and anomaly detection as underscored in latest literature~\cite{almasan2022network}. The architectural NDT and Physical Twin (PT) interaction as illustrated in Figure~\ref{fig:ndt_pt} typically consists of data acquisition, a digital replica and a feedback loop to the physical network. Underlying algorithms such as artificial intelligence (AI) models are trained and deployed to emulate network dynamics. Recent surveys such as \cite{Mihai}, and \cite{Wu} emphasize NDTs as key enablers for next-generation networks (NGN), highlighting open research issues such as real-time synchronization, scalability and high-fidelity modeling.
Despite the efficient network operation NDTs provide, efficient operation poses significant challenges especially in maintaining synchronization between physical and digital replica, which requires continuous exchange of network state information, thereby being costly in terms of bandwidth and energy. In vehicular digital twin networks, the combination of intra‑twin and inter‑twin communications leads to increased response times, managing communication overhead on the cloud layer becomes essential as the authors in~\cite{VDT_Priority} noted. Several approaches have been proposed to mitigate these challenges, such as priority-based communication schemes that reduce latency by categorizing messages according to their urgency as discussed in~\cite{VDT_Priority}, and edge‑computing frameworks that keep processing operations locally at the edge, so as to reduce the amount of data sent to the cloud. Additionally, as presented in recent work from \cite{sengendo2025building}, adaptive control mechanisms such as PID controllers have been proposed to dynamically improve real-time synchronization between digital twins and physical networks. 

Exploring more within in the existing state-of-the-art, recent research on AI-enabled Network Digital Twins (NDTs) indicates that the primary role of artificial intelligence is to enhance prediction accuracy, state estimation, and synchronization fidelity between the physical and digital domains rather than to directly minimize communication overhead. Machine learning (ML) techniques, including deep learning, graph neural networks, and reinforcement learning (RL), are predominantly employed for traffic forecasting, network behavior prediction, anomaly detection, and closed-loop control, enabling the digital twin to anticipate future network states and support proactive decision-making. 
Works presented in \cite{almasan2022network} additionally emphasize that modern ML forms a core component of NDT architectures by enabling accurate real-time network modeling and performance prediction. Similarly, in \cite{shin2023network} the authors demonstrate the use of AI-driven traffic prediction models within digital twin architectures to forecast future network conditions and improve operational efficiency. 

Recent surveys, for example, in \cite{sanjalawe2026bridging} further discuss synchronization and predictive intelligence as fundamental functions of AI-driven digital twins, with AI supporting model creation, synchronization, prediction, and feedback control throughout the digital twin lifecycle. Although communication overhead is recognized as a significant challenge in maintaining synchronization between physical and virtual entities, it is generally treated as a system-level constraint rather than the primary optimization objective of AI models. Consequently, current AI-based NDT research largely focuses on achieving accurate prediction and maintaining high-fidelity synchronization, while communication overhead reduction is typically less explored.

\subsection{Topological Signal Processing}

Graph signal processing (GSP) leverages classical signal processing to data residing on the nodes of a graph~\cite{GSP_Overview}. It provides tools for filtering, sampling, reconstruction and spectral analysis of signals defined on irregular graph domains. For instance, work presented in \cite{polverini2024reducing} applies graph-structured assumptions in network telemetry by introducing SPAN, which reduces overhead through spatial sampling and reconstruction over the network topology \cite{polverini2024reducing}. This is achieved by exploiting spatial correlations between telemetry measurements to infer missing network state information with high accuracy. However, graphs capture only pairwise interactions and many systems exhibit higher‑order relations. Topological signal processing (TSP) enhances GSP by considering signals on nodes, edges, and higher‑dimensional simplices~\cite{TSP_Simplicial}. The TSP framework derives sampling theory and inference algorithms for higher‑order signals and has been applied to flow data on edges of graphs and other multi‑way interactions. One limitation of simplicial complexes is the inclusion property. If a simplex belongs to the complex, all its faces must also belong. In many applications, this constraint is unnecessary and restrictive. Authors in \cite{TSP_Cell} show that cell complexes provide a natural hierarchical structure and lead to better sparsity accuracy trade-offs than graphs or simplicial complexes~\cite{TSP_Cell}. Topological deep learning further extends these ideas to neural network architectures. Survey articles on message-passing topological neural networks underscore that graphs are limited to pairwise relations and advocate the use of simplicial, cellular, and combinatorial complexes to capture higher‑order interactions~\cite{TDL_Survey}.

Additionally, in line with addressing the communication overhead bottleneck in digital twin systems, federated learning approaches have been explored for digital twins that train models locally and exchange only model parameters, thus reducing data transfer; hierarchical federated learning schemes further compress updates and improve communication efficiency \cite{9145588}. Although effective, minimizing communication overhead during network monitoring is still often overlooked, particularly when applying GSP and TSP in NTDs. This is critical for efficient bandwidth utilization, especially when monitoring data must be transmitted and replicated in a digital twin. Addressing this remains an open research gap that we explore. In the next section, we present the methodology.

\section{Methodology}
\label{sec:methodology}

We develop the ST-NDT framework in three stages: cell-complex representation of the network,
Hodge-spectral sensor placement, and Tikhonov signal reconstruction.
The full pipeline is outlined in Algorithms~\ref{alg:placement}--\ref{alg:recon}.

\subsection{Network Representation as a Cell Complex}
\label{sec:cell_complex}

Let $\mathcal{G} = (\mathcal{V}, \mathcal{E})$ be a connected, undirected network with
$N = |\mathcal{V}|$ nodes and $E = |\mathcal{E}|$ edges.
We embed $\mathcal{G}$ in a \emph{cell complex} $\mathcal{X}$~\cite{TSP_Cell} by augmenting it with $P$ minimal
cycles (polygons) identified via the minimum cycle basis, and encode the
topology via two boundary operators:
\begin{align}
  \mathbf{B}_1 &\in \mathbb{R}^{N \times E}, &
  \mathbf{B}_2 &\in \mathbb{R}^{E \times P},
\end{align}
where $[\mathbf{B}_1]_{v,e} \in \{-1,0,+1\}$ is oriented node--edge incidence and
$[\mathbf{B}_2]_{e,p} \in \{-1,0,+1\}$ is oriented edge--polygon incidence.
By construction, $\mathbf{B}_1\mathbf{B}_2 = \mathbf{0}$~\cite{TSP_Simplicial,TSP_Cell}.
The \emph{total edge Laplacian} is
\begin{equation}
  \mathbf{L}_1 = \underbrace{\mathbf{B}_1^\top\mathbf{B}_1}_{\mathbf{L}_1^d}
               + \underbrace{\mathbf{B}_2\mathbf{B}_2^\top}_{\mathbf{L}_1^u}
               \;\in\mathbb{R}^{E\times E},
\end{equation}
whose eigendecomposition $\mathbf{L}_1 = \mathbf{U}\boldsymbol{\Lambda}\mathbf{U}^\top$
yields an orthonormal basis $\mathbf{U}$ with eigenvalues
$0 \le \lambda_1 \le \cdots \le \lambda_E$~\cite{TSP_Simplicial,TSP_Cell}.

\subsection{Hodge Decomposition of Edge Signals}
\label{sec:hodge}

Every edge signal $\mathbf{f} \in \mathbb{R}^E$ admits the unique orthogonal
splitting~\cite{TSP_Simplicial}:
\begin{equation}
  \mathbf{f} = \mathbf{f}_{\mathrm{irr}} + \mathbf{f}_{\mathrm{sol}} + \mathbf{f}_{\mathrm{harm}},
  \label{eq:hodge}
\end{equation}
where $\mathbf{f}_{\mathrm{irr}} \in \mathrm{im}(\mathbf{B}_1^\top)$ captures gradient
(routing-metric) flows, $\mathbf{f}_{\mathrm{sol}} \in \mathrm{im}(\mathbf{B}_2)$ captures
circular cycle flows, and $\mathbf{f}_{\mathrm{harm}} \in \ker(\mathbf{L}_1)$ captures
persistent flows through topological holes ($\dim = \beta_1 = E - N + \beta_0$)~\cite{TSP_Simplicial}.

\subsection{Problem Formulation}
\label{sec:problem}

Let $S \subset \mathcal{E}$, $|S| = m \ll E$, be the monitored edge set. Given
$\mathbf{y}_S = \mathbf{f}[S] + \boldsymbol{\eta}$, the sparse monitoring problem is to recover
$\hat{\mathbf{f}} \approx \mathbf{f}$ by jointly optimising (i) \textbf{sensor placement}: select
$S$ to maximise information about $\mathbf{f}$, and (ii) \textbf{reconstruction}: estimate
$\hat{\mathbf{f}}$ from $\mathbf{y}_S$ and the network topology.

\subsection{Adaptive Spectral Bandlimit}
\label{sec:bandlimit}

Hodge-spanning bandlimit:
Given sensor budget $m$ and harmonic estimate
$\hat{\beta}_1 = \#\{i : \lambda_i < 10^{-2}\}$, the \emph{Hodge-spanning bandlimit} is
\begin{equation}
  k^* = \min\!\bigl(\max(\hat{\beta}_1 + m,\; 2m),\; E\bigr).
  \label{eq:kstar}
\end{equation}

The condition $k^* \ge \hat{\beta}_1 + m$ ensures the bandlimited basis
$\mathbf{U}_{k^*} = \mathbf{U}[:,1{:}k^*]$ spans the harmonic, solenoidal, and a portion of the
irrotational subspace simultaneously, so the placement algorithm covers all Hodge components.
See steps~1--3 of Algorithm~\ref{alg:placement}.

\subsection{Topological Sensor Placement}
\label{sec:placement_alg}

Given $\mathbf{U}_{k^*}$, the optimal set $S$ maximises
$\sigma_{\min}(\mathbf{U}_{k^*}[S,:])$, which is approximated by QR column
pivoting, as summarised in Algorithm~\ref{alg:placement}.
The initial QR-pivot selection is subsequently refined via a single-pass greedy
conditioning swap (See Algorithm~\ref{alg:refine}).

\begin{algorithm}[htbp]
\SetAlgoLined
\DontPrintSemicolon
\KwIn{Graph $\mathcal{G}$, sensor budget $m$, edge Laplacian $\mathbf{L}_1$}
\KwOut{Sensor edge set $S$}
\BlankLine
Compute $(\boldsymbol{\Lambda}, \mathbf{U}) \leftarrow \mathrm{eigh}(\mathbf{L}_1 + \epsilon\mathbf{I})$\;
Estimate $\hat{\beta}_1 \leftarrow \#\{i : \lambda_i < 10^{-2}\}$\;
$k^* \leftarrow \min(\max(\hat{\beta}_1 + m,\,2m),\,E)$;\quad
$\mathbf{U}_{k^*} \leftarrow \mathbf{U}[:,1{:}k^*]$\;
$[\mathbf{Q},\,\mathbf{R},\,\boldsymbol{\pi}] \leftarrow \mathrm{QR}(\mathbf{U}_{k^*}^\top,\;\text{pivot}=\mathrm{True})$\;
$S \leftarrow \boldsymbol{\pi}[1{:}m]$\tcp*{first $m$ pivot indices}
\If{$|S| < m$}{
  supplement from top leverage-score edges\;
}
$S \leftarrow \mathrm{GreedyRefine}(S, \mathbf{U}_{k^*}, m)$\tcp*{single-pass swap; see Alg.~\ref{alg:refine}}
\KwRet{$S$}
\caption{Topological Sensor Placement}
\label{alg:placement}
\end{algorithm}

\begin{algorithm}[htbp]
\SetAlgoLined
\DontPrintSemicolon
\KwIn{Initial set $S_0$, basis $\mathbf{U}_k$, budget $m$}
\KwOut{Refined set $S$}
\BlankLine
\If{$|S_0| < k$}{\KwRet{$S_0$}}
$S \leftarrow S_0$;\quad
$\mathcal{C} \leftarrow \mathrm{top}$-$30$ edges by $\|\mathbf{U}_k[e,:]\|_2$, $e \notin S$\;
$\sigma^* \leftarrow \sigma_{\min}(\mathbf{U}_k[S,:])$\;
\For{$j \in \mathrm{argsort}(\|\mathbf{U}_k[S,:]\|_{\mathrm{row}})[:5]$}{
  \For{$e \in \mathcal{C}$}{
    $S' \leftarrow S$;\; $S'[j] \leftarrow e$\;
    \If{$\sigma_{\min}(\mathbf{U}_k[S',:]) > 1.02\,\sigma^*$}{
      $S \leftarrow S'$;\; $\sigma^* \leftarrow \sigma_{\min}(\mathbf{U}_k[S,:])$;\; \textbf{break}\;
    }
  }
}
\KwRet{$S$}
\caption{Greedy Conditioning Refinement}
\label{alg:refine}
\end{algorithm}

\subsection{Sparse Signal Reconstruction}
\label{sec:reconstruction}

Given measurements $\mathbf{y}_S$, we recover the full signal via Tikhonov regularization
with $\mathbf{L}_1$ as the smoothness prior, as summarized in
Algorithm~\ref{alg:recon}:
\begin{equation}
  \hat{\mathbf{f}} = \argmin_{\mathbf{f}}
    \bigl\|\mathbf{P}_S(\mathbf{f} - \mathbf{y}_{\mathrm{full}})\bigr\|_2^2
    + \lambda\,\|\mathbf{L}_1^{1/2}\mathbf{f}\|_2^2,
  \label{eq:tikhonov}
\end{equation}
whose closed-form solution satisfies
$(\mathbf{P}_S + \lambda\,\|\mathbf{L}_1\|_2\,\mathbf{L}_1)\,\hat{\mathbf{f}}
= \mathbf{P}_S\,\mathbf{y}_{\mathrm{full}}$.
The regularization parameter is tuned adaptively as shown in Equation \eqref{eq:lambda}:
\begin{equation}
  \lambda = \mathrm{clip}\!\left(
    \frac{5 \times 10^{-4}}{\max(1,\sqrt{\widehat{\mathrm{SNR}}})},\;
    10^{-5},\; 10^{-2}\right),
  \label{eq:lambda}
\end{equation}
where $\widehat{\mathrm{SNR}}$ is estimated from the observed measurements via a MAD-based
noise estimate. Because $\mathbf{L}_1 = \mathbf{L}_1^d + \mathbf{L}_1^u$, the regularizer
simultaneously penalizes irrotational and solenoidal roughness; topological sensors that
observe solenoidal cycle edges allow the solver to propagate those flows accurately, while
hub-centric baselines that miss cycle edges incur systematic reconstruction errors.

\begin{algorithm}[htbp]
\SetAlgoLined
\DontPrintSemicolon
\KwIn{Measurements $\mathbf{y}_S$, placement $S$, Laplacian $\mathbf{L}_1$}
\KwOut{Reconstructed signal $\hat{\mathbf{f}} \in \mathbb{R}^E$}
\BlankLine
$\mathbf{y}_{\mathrm{full}}[S] \leftarrow \mathbf{y}_S$;\quad
$\mathbf{P}_S \leftarrow \mathrm{diag}(\mathbf{1}_S)$\;
Compute $\lambda$ via~\eqref{eq:lambda}\;
$\hat{\mathbf{f}} \leftarrow (\mathbf{P}_S + \lambda\,\|\mathbf{L}_1\|_2\,\mathbf{L}_1)^{-1}
  \mathbf{P}_S\,\mathbf{y}_{\mathrm{full}}$\;
$\hat{\mathbf{f}} \leftarrow \max(\hat{\mathbf{f}},\,\mathbf{0})$;\quad
$\hat{\mathbf{f}}[S] \leftarrow \mathbf{y}_S$\;
$\hat{\mathbf{f}} \leftarrow \mathrm{GraphSmooth}(\hat{\mathbf{f}},S,\alpha{=}0.2,\text{iters}{=}3)$\;
\KwRet{$\hat{\mathbf{f}}$}
\caption{Sparse Reconstruction via Topology-Informed Tikhonov}
\label{alg:recon}
\end{algorithm}

\paragraph{Computational complexity.}
Eigendecomposition of $\mathbf{L}_1$ costs $\mathcal{O}(E^3)$; QR pivoting costs
$\mathcal{O}(k^* E^2)$; the Tikhonov solve costs $\mathcal{O}(E^3)$.
For the Cogentco network ($E = 243$) the full pipeline completes fast. In the next section, we detail our experimental setup.

\section{Experimental Setup}
\label{sec:experimental_setup}

\subsection{Dataset: Cogentco ISP Backbone Network}
\label{subsec:dataset}

\begin{figure}[h]
\centering
\includegraphics[width=0.80\linewidth]{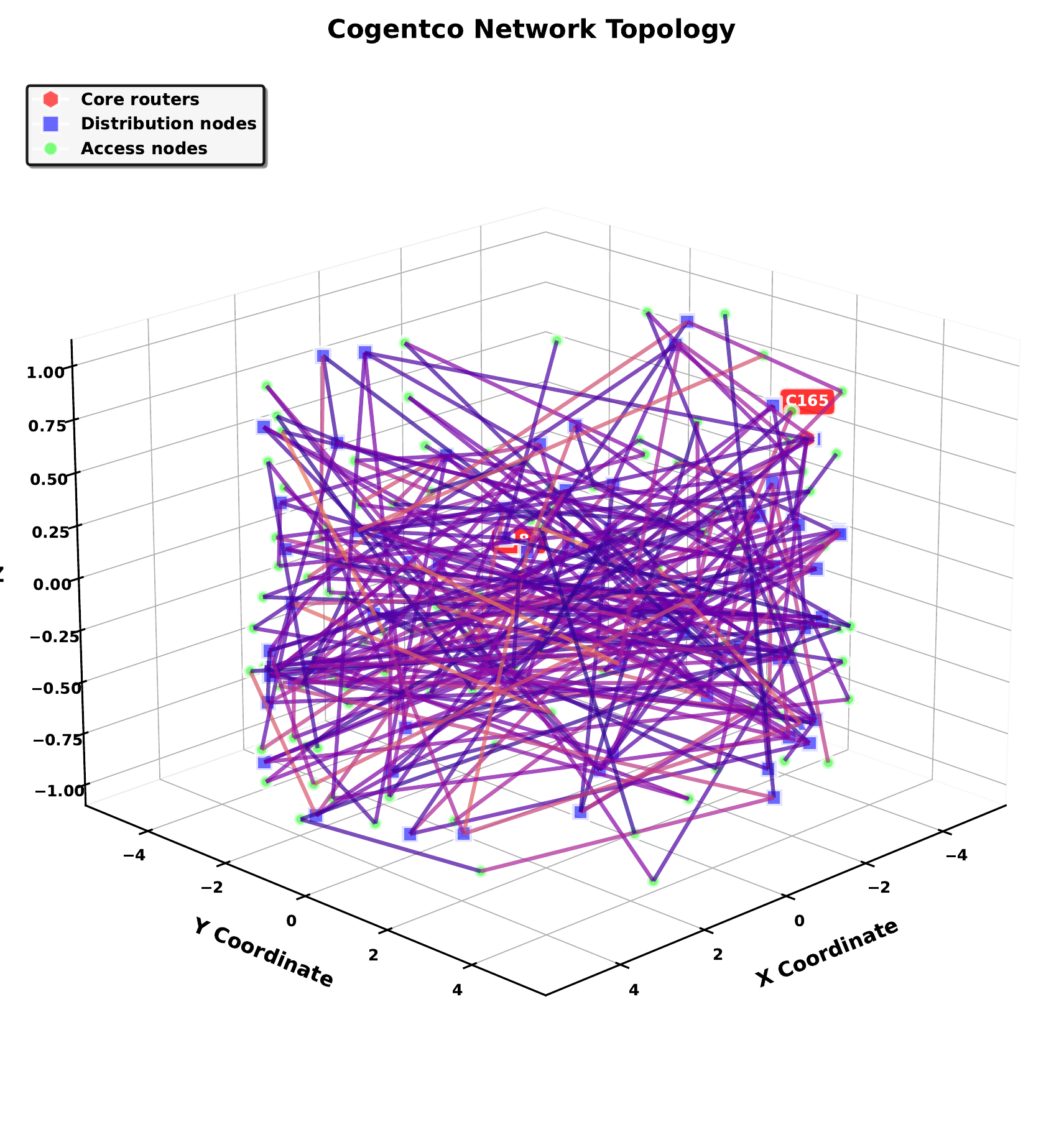}
\caption{Cogentco network. Nodes coloured by hierarchy
  (red = core, blue = distribution, green = access); edges by traffic
  intensity (purple--magenta). The dense mesh structure drives topology-aware placement.}
\label{fig:network_topology}
\end{figure}

We evaluate ST-NDT on the Cogentco backbone network from the Internet Topology
Zoo~\cite{3D-internet-zoo}, preprocessed to $|\mathcal{V}| = 197$ nodes and
$|\mathcal{E}| = 243$ edges ($\bar{d} \approx 2.46$). Key topological invariants are
listed in Table~\ref{tab:network_stats} and a 3D visualization shown in Figure~\ref{fig:network_topology}; the $\beta_1 = 39$ independent cycles drive
the sensor placement strategy.

\begin{table}[t]
\centering
\caption{Cogentco network topology statistics}
\label{tab:network_stats}

\setlength{\tabcolsep}{10pt} 
\renewcommand{\arraystretch}{1.1} 

\begin{tabular}{lc}
\toprule
\textbf{Property} & \textbf{Value} \\
\midrule
Nodes $|\mathcal{V}|$ & 197 \\
Edges $|\mathcal{E}|$ & 243 \\
Identified polygons $|\mathcal{F}|$ & 8 \\
Connected components $\beta_0$ & 1 \\
Topological holes (cycles) $\beta_1$ & 39 \\
Average node degree $\bar{d}$ & 2.46 \\
\bottomrule
\end{tabular}

\end{table}

\subsection{Traffic Signal Generation}
\label{subsec:signal_generation}

We construct a composite edge signal with both irrotational and solenoidal components.

\paragraph{Irrotational component.}
A node potential $\phi \in \mathbb{R}^{|\mathcal{V}|}$ is drawn from a uniform
distribution and projected onto edges:
\begin{equation}
  f_{\mathrm{irr}} = |B_1^{\top} \phi|^{0.85},
\end{equation}
where the sub-linear exponent yields a realistic heavy-tailed traffic distribution.

\paragraph{Solenoidal component.}
Cycle amplitudes $\mathbf{a} \in \mathbb{R}^{|\mathcal{F}|}$ are assigned linearly in $[1,3]$~\cite{TSP_Simplicial}:
\begin{equation}
  f_{\mathrm{sol}} = |B_2\, \mathbf{a}|.
\end{equation}
The components are mixed with a normalised weight:
\begin{equation}
  f = f_{\mathrm{irr}} + \alpha_s\, f_{\mathrm{sol}}, \quad
  \alpha_s = \frac{\overline{f_{\mathrm{irr}}}}{\max(f_{\mathrm{sol}}) + \epsilon},
\end{equation}
ensuring both Hodge components contribute at comparable scale.

\subsection{Baseline Sensor Placement Methods}
\label{subsec:baselines}

We compare ST-NDT against four classical approaches. The degree-based method selects edges incident to the highest-degree nodes, while the betweenness centrality method selects the top-$m$ edges according to their edge betweenness scores \cite{basuchowdhuri2016detecting}. The core-number ($k$-core) method selects edges whose endpoints belong to the highest $k$-core, whereas the random method selects $m$ edges uniformly at random using a fixed seed for reproducibility \cite{7511156}. Table~\ref{tab:method_comparison} summarises the structural properties of all five methods.

\begin{table*}[t]
\centering
\caption{Structural comparison of sensor placement strategies.}
\label{tab:method_comparison}

\setlength{\tabcolsep}{4.5pt}
\renewcommand{\arraystretch}{1.1}

\begin{tabular}{lcccc}
\toprule
\textbf{Method} & \textbf{Topology aware} & \textbf{Selection criterion} & \textbf{Deterministic} & \textbf{Global coverage} \\
\midrule
\textbf{ST-NDT (Topological)} & \textbf{Yes} & \textbf{Hodge eigenbasis (QR pivot)} & \textbf{Yes} & \textbf{Yes} \\
Degree-based           & No & Node degree rank       & Yes & No \\
Betweenness centrality & No & Edge betweenness score & Yes & Partial \\
Core-number            & No & $k$-core decomposition & Yes & No \\
Random                 & No & Uniform random         & No  & Random \\
\bottomrule
\end{tabular}

\end{table*}

\subsection{Evaluation Metrics}
\label{subsec:metrics}

Three complementary metrics were used to assess reconstruction fidelity.
Let $f, \hat{f} \in \mathbb{R}^{|\mathcal{E}|}$ be the ground-truth and
reconstructed edge signals respectively.

\paragraph{Normalized Reconstruction Error (NRE).}
NRE measuring the relative reconstruction error normalized by the signal energy,
making it scale-invariant and comparable across different traffic conditions:
\begin{equation}
  \mathrm{NRE} = \frac{\|f - \hat{f}\|_2}{\|f\|_2}.
\end{equation}
With the formulation as shown above, a value of $0$ indicates perfect reconstruction, $1$ means the error is equal in magnitude to the signal itself.

\paragraph{Mean Absolute Error (MAE).}
MAE quantifying the average per-edge deviation between the true and reconstructed
traffic flows, expressed in the original traffic units as in Equation \eqref{eq:mae}:
\begin{equation}
  \mathrm{MAE} = \frac{1}{|\mathcal{E}|} \sum_{e} |f_e - \hat{f}_e|.
  \label{eq:mae}
\end{equation}
It provides an interpretable measure of how far the reconstruction deviates
from ground truth on a typical edge.

\paragraph{Root Mean Square Error (RMSE).}
RMSE penalises large per-edge errors more heavily than MAE by squaring the
residuals before averaging \cite{chicco2021coefficient}, making it sensitive to edges with poor reconstruction see Equation \eqref{eq:rmse}:
\begin{equation}
  \mathrm{RMSE} = \sqrt{\frac{1}{|\mathcal{E}|} \sum_{e} (f_e - \hat{f}_e)^2}.
  \label{eq:rmse}
\end{equation}
Together, MAE and RMSE provide a complete picture of both typical and
worst-case reconstruction behaviour across the network.

The monitoring budget $m$ is varied across five levels $10\%$, $15\%$, $20\%$, $30\%$,
and $40\%$ of $|\mathcal{E}|$ (i.e., 24, 36, 48, 72, and 97 sensor edges). At each level,
each method places $m$ sensors, unobserved edges are set to zero, and Tikhonov reconstruction
recovers the full signal; errors are evaluated on all $|\mathcal{E}|$ edges.
All methods are deterministic; the random baseline uses a budget-specific fixed seed. All methods where implemented in Python~3.13.

\section{Results and Discussion}
\label{sec:results}

This section presents the empirical results from the proposed framework benchmarked against four baseline sensor placement strategies. Results are categorized into four presentations: (i) per-method reconstruction accuracy at 20\% monitoring budget; (ii) error distribution characteristics across all edges;
(iii) the spatial structure of sensor placements; and (iv) performance across varying budget
levels.

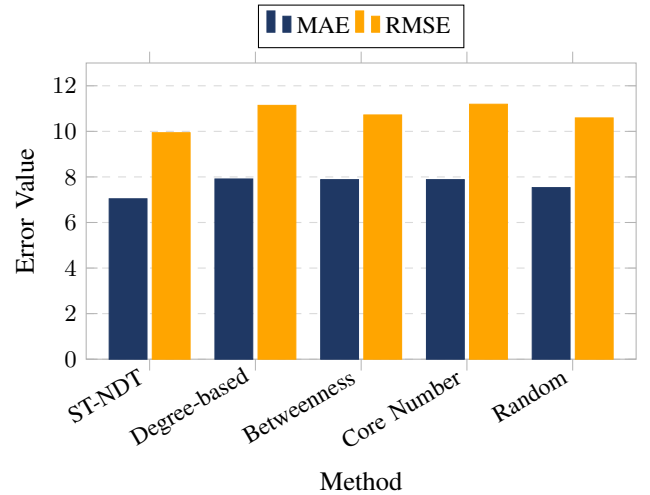
\begin{figure}[h]
  \centering
  \begin{tikzpicture}
  \begin{axis}[
      ybar,
      bar width=0.5cm,
      width=\columnwidth,
      height=5.5cm,
      enlarge x limits=0.15,
      ylabel={Error Value},
      xlabel={Method},
      symbolic x coords={ST-NDT, Degree-based, Betweenness, {Core Number}, Random},
      xtick=data,
      x tick label style={
          rotate=30,
          anchor=east,
          font=\small
      },
      ymin=0, ymax=13,
      ytick={0,2,4,6,8,10,12},
      yticklabel style={/pgf/number format/fixed, /pgf/number format/precision=1},
      legend style={
          at={(0.5,1.05)},
          anchor=south,
          legend columns=2,
          font=\small,
          draw=black,
          fill=white,
      },
      ymajorgrids=true,
      grid style={dashed, gray!30},
      axis line style={gray!60},
      tick style={gray!60},
  ]

  \addplot[fill=navyblue, draw=navyblue] coordinates {
      (ST-NDT,   7.04)
      (Degree-based,  7.91)
      (Betweenness,   7.88)
      ({Core Number}, 7.88)
      (Random,        7.53)
  };
  \addlegendentry{MAE}

  \addplot[fill=orange, draw=orange] coordinates {
      (ST-NDT,   9.94)
      (Degree-based,  11.14)
      (Betweenness,   10.72)
      ({Core Number}, 11.19)
      (Random,        10.59)
  };
  \addlegendentry{RMSE}

  \end{axis}
  \end{tikzpicture}
  \caption{MAE and RMSE at 20\% monitoring budget.
  ST-NDT achieving the lowest error on both metrics.}
  \label{fig:mae_rmse_comparison}
\end{figure}
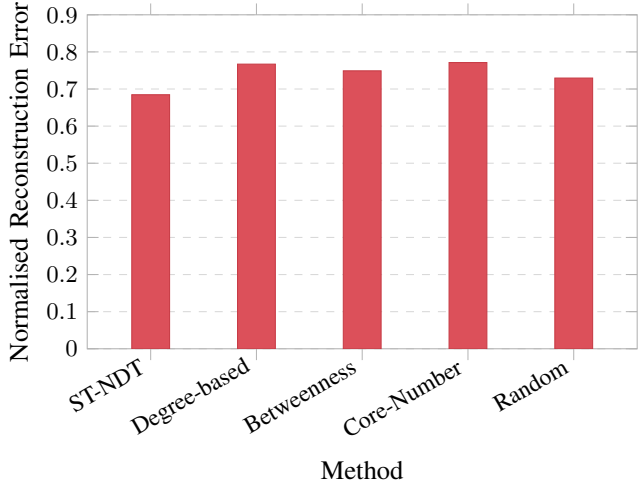
\begin{figure}[h]
  \centering
  \begin{tikzpicture}
  \begin{axis}[
      ybar,
      bar width=0.5cm,
      width=\columnwidth,
      height=6cm,
      enlarge x limits=0.15,
      ylabel={Normalised Reconstruction Error},
      xlabel={Method},
      symbolic x coords={ST-NDT, Degree-based, Betweenness, {Core-Number}, Random},
      xtick=data,
      x tick label style={
          rotate=30,
          anchor=east,
          font=\small
      },
      ymin=0, ymax=0.9,
      ytick={0.0,0.1,0.2,0.3,0.4,0.5,0.6,0.7,0.8,0.9},
      yticklabel style={/pgf/number format/fixed, /pgf/number format/precision=1},
      ymajorgrids=true,
      grid style={dashed, gray!30},
      axis line style={gray!60},
      tick style={gray!60},
      cycle list={
          {fill={rgb,255:red,220;green,80;blue,90},   draw={rgb,255:red,200;green,60;blue,70}},
          {fill={rgb,255:red,80;green,100;blue,200},  draw={rgb,255:red,60;green,80;blue,180}},
          {fill={rgb,255:red,50;green,140;blue,130},  draw={rgb,255:red,30;green,120;blue,110}},
          {fill={rgb,255:red,140;green,90;blue,200},  draw={rgb,255:red,120;green,70;blue,180}},
          {fill={rgb,255:red,170;green,140;blue,220}, draw={rgb,255:red,150;green,120;blue,200}},
      },
  ]
  \addplot coordinates {
      (ST-NDT,   0.6846)
      (Degree-based,  0.7670)
      (Betweenness,   0.7490)
      ({Core-Number}, 0.7713)
      (Random,        0.7295)
  };
  \end{axis}
  \end{tikzpicture}
  \caption{Normalised reconstruction error ($\|f - \hat{f}\|_2 / \|f\|_2$) at 20\% monitoring budget.}
  \label{fig:nre}
\end{figure}
\subsection{Reconstruction Accuracy at 20\% Budget}
\label{subsec:recon_accuracy}

Figure~\ref{fig:mae_rmse_comparison} depicts the Mean Absolute Error (MAE) and Root Mean Square
Error (RMSE) for each method at the 20\% monitoring budget ($m = 48$ sensors on 243 edges). In Table~\ref{tab:accuracy_20}, we additionally report the NRE achieved by all five frameworks under a 20\% monitoring budget. Alongside the absolute reconstruction error difference ($\Delta$NRE) relative to the proposed ST-NDT topological approach, we also present the corresponding percentage improvement obtained by ST-NDT against each baseline method, computed as shown in Equation \eqref{eq:percentage_improvement}.

\begin{equation}
\text{Improvement (\%)} =
\frac{\text{NRE}_{\text{baseline}} - \text{NRE}_{\text{ST-NDT}}}
{\text{NRE}_{\text{baseline}}} \times 100
\label{eq:percentage_improvement}
\end{equation}

As illustrated in Figure~\ref{fig:nre}, the proposed topological method consistently achieves the lowest normalised reconstruction error across all baselines, confirming the trend reported in Table~\ref{tab:accuracy_20}. In particular, ST-NDT attains an NRE of $0.6846$, outperforming the degree-based baseline ($0.7679$) by $10.85\%$, betweenness centrality ($0.7390$) by $7.36\%$, core-number placement ($0.7713$) by $11.24\%$, and the fixed-seed random baseline ($0.7295$) by $6.15\%$. These improvements are further aligned with in both the MAE and RMSE results depicted in Figure~\ref{fig:mae_rmse_comparison}, where ST-NDT outperforms the rest in all performance metrics, indicating that the topological framework reduces reconstruction error more consistently across the network.

\begin{table}[t]
\centering
\caption{Normalised Reconstruction Error (NRE) comparison of different frameworks.}
\label{tab:accuracy_20}

\begin{tabular*}{\columnwidth}{@{\extracolsep{\fill}}lccc}
\toprule
\textbf{Method} & \textbf{NRE} $\downarrow$ & $\boldsymbol{\Delta}$\textbf{NRE} & \textbf{Improvement (\%)} \\
\midrule
\textbf{ST-NDT (Topological)} & \textbf{0.6846} & ---     & --- \\
Betweenness centrality        & 0.7390          & 0.0544 & 7.36\% \\
Degree-based                  & 0.7679          & 0.0833 & 10.85\% \\
Core-number                   & 0.7713          & 0.0867 & 11.24\% \\
Random                        & 0.7295          & 0.0449 & 6.15\% \\
\bottomrule
\end{tabular*}

\end{table}

The margin over degree-based and core-number methods is particularly minimal as both mechanisms
favor hub edges incident to high-connectivity nodes, but the clustering leaves peripheral edges which carry real traffic entirely unobserved. In contrast, the Hodge-spectral placement mechanism selects edges that span the dominant eigenvectors of the graph Laplacian $L_1$,
implicitly achieving spatial coverage across the entire network.
Expanding more on the analysis presented in Table~\ref{tab:accuracy_20}, Figure~\ref{fig:nre} summarizes the normalized reconstruction error plots for all methods at the
20\% budget, providing a plot view of relative reconstruction quality performance. The consistent gap between the topological curve and all baselines
demonstrates that the gain arises from sensor \emph{placement} quality, not reconstruction algorithm differences as all methods share the same Tikhonov solver.

\subsection{Error Distribution Analysis}
\label{subsec:error_dist}

\begin{figure}[h]
  \centering
  \begin{tikzpicture}
  \begin{axis}[
      width=\columnwidth,
      height=7cm,
      boxplot/draw direction=y,
      ylabel={Relative Error (\%)},
      xlabel={Method},
      xtick={1,2,3,4,5},
      xticklabels={ST-NDT, Degree-based, Betweenness, {Core-Number}, Random},
      x tick label style={
          rotate=30,
          anchor=east,
          font=\small
      },
      ymin=0, ymax=150,
      ytick={0,25,50,75,100,125,150},
      ymajorgrids=true,
      grid style={dashed, gray!30},
      axis line style={gray!60},
      tick style={gray!60},
      boxplot/every whisker/.style={solid, black, line width=0.8pt},
      boxplot/every median/.style={solid, black, line width=1.2pt},
      boxplot/every box/.style={solid, draw=black, line width=0.8pt},
      boxplot/whisker extend=0.25,
  ]

  \addplot+[
      boxplot prepared={
          lower whisker=2,
          lower quartile=11.48,
          median=55.54,
          upper quartile=78.99,
          upper whisker=112,
      },
      fill=blue!25, draw=black,
  ] coordinates {};
  \addplot+[only marks, mark=o, mark size=1.5pt, draw=black]
      coordinates {(1,135) (1,140)};

  \addplot+[
      boxplot prepared={
          lower whisker=2,
          lower quartile=18.64,
          median=63.46,
          upper quartile=86.70,
          upper whisker=120,
      },
      fill=blue!25, draw=black,
  ] coordinates {};
  \addplot+[only marks, mark=o, mark size=1.5pt, draw=black]
      coordinates {(2,138) (2,143)};

  \addplot+[
      boxplot prepared={
          lower whisker=2,
          lower quartile=17.23,
          median=66.09,
          upper quartile=87.64,
          upper whisker=122,
      },
      fill=blue!25, draw=black,
  ] coordinates {};
  \addplot+[only marks, mark=o, mark size=1.5pt, draw=black]
      coordinates {(3,136) (3,141)};

  \addplot+[
      boxplot prepared={
          lower whisker=2,
          lower quartile=15.35,
          median=63.67,
          upper quartile=85.87,
          upper whisker=121,
      },
      fill=blue!25, draw=black,
  ] coordinates {};
  \addplot+[only marks, mark=o, mark size=1.5pt, draw=black]
      coordinates {(4,137) (4,142)};

  \addplot+[
      boxplot prepared={
          lower whisker=2,
          lower quartile=15.11,
          median=60.83,
          upper quartile=82.78,
          upper whisker=118,
      },
      fill=blue!25, draw=black,
  ] coordinates {};
  \addplot+[only marks, mark=o, mark size=1.5pt, draw=black]
      coordinates {(5,134) (5,139)};

  \end{axis}
  \end{tikzpicture}
  \caption{Distribution of per-edge relative reconstruction errors (\%) for each method at 20\% monitoring budget. ST-NDT achieving the lowest median error.}
  \label{fig:boxplot}
\end{figure}
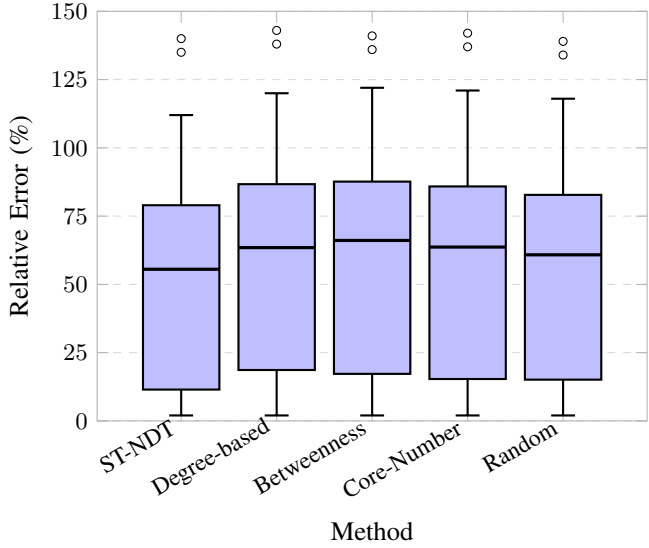
In Figure~\ref{fig:boxplot}, we show the distribution of per-edge relative reconstruction errors (\%) for each method at a 20\% monitoring budget. ST-NDT (Topological) achieved the lowest median error of $\approx$55.54\%, with an IQR of 67.50\% (Q1\,=\,11.48\%, Q3\,=\,78.99\%), reflecting a comparatively compact and consistent error distribution. The Degree-based and Core-Number baselines exhibited higher medians of $\approx$ 63.46\% and 63.67\% respectively, with IQRs around 68--71\%, indicating greater reconstruction variability across edges. Betweenness centrality yielded the higher median error at $\approx$ 66.09\% attributed to the large number of edges that are completely unobserved, despite it having a lower mean. The Random baseline, despite its simplicity, performs comparably to the informed baselines with a median of $\approx$ 60.83\% and an IQR of $\approx$ 67.67\%. All methods display right-skewed distributions with outliers extending well beyond the $1.5\times$IQR whisker range, highlighting the difficulty of reconstructing a small number of structurally critical or heavily loaded edges. These results further reinforce that topology-aware edge selection consistently yields lower and more stable relative errors than uninformed or degree-only alternatives.

\subsection{Sensor Placement Visualisation}
\label{subsec:sensor_map}

\begin{figure*}[ht]
\centering
\includegraphics[width=0.95\linewidth]{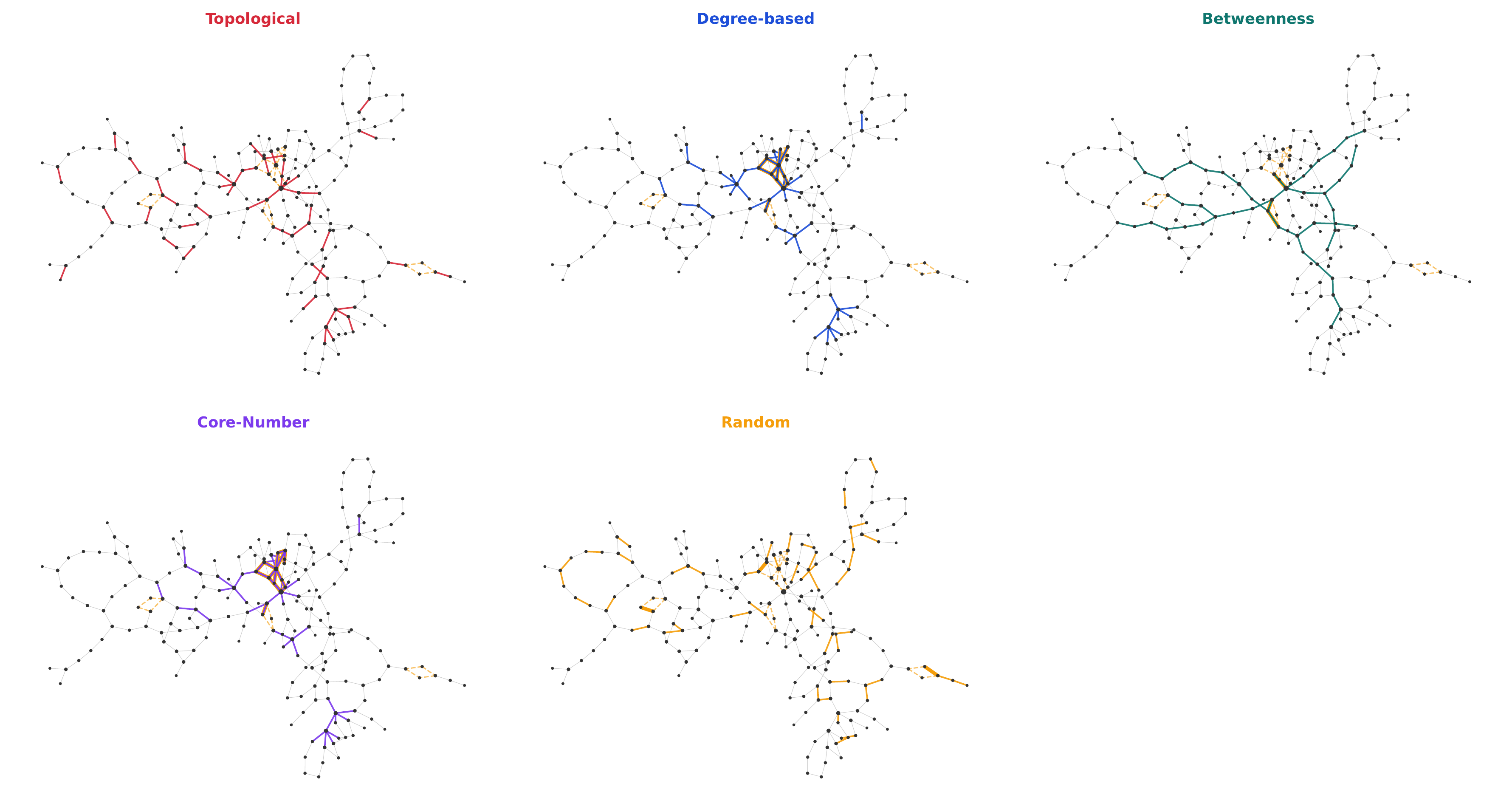}
\caption{Sensor edge selection at 20\% budget (48 of 243 edges) on the Cogentco network topology; gold edges mark polygon boundaries. ST-NDT (Topological) spreads sensors network-wide; hub-centric methods (degree, core) leave peripheral edges unmonitored.}
\label{fig:sensor_map}
\end{figure*}

In Figure~\ref{fig:sensor_map}, we provide a visualization of the 48 sensor edges selected by each method, overlaid on a force-directed layout of the Cogentco network. The sub-panels in Figure~\ref{fig:sensor_map} each showing the same network layout but with
different sets of 48 sensor edges highlighted. Key observations noted:

\begin{enumerate}
  \item \textbf{ST-NDT(Topological):} Sensor edges (colored red) are distributed across the entire network
    graph, covering both the dense hub region and sparse peripheral branches. This spatial spread
    is an emergent property of the QR column-pivoting criterion: selecting edges that collectively
    span the dominant eigenvectors of $L_1$ forces geographic diversity, because different
    eigenvectors have support on different parts of the network. The gold polygon edges are also
    well represented, as the adaptive bandlimit $k^* \geq \beta_1 + m$ ensures that the harmonic
    subspace (topological holes) is captured.

  \item \textbf{Degree-based:} Sensors cluster near the hub nodes in the centre and eastern
    portion of the layout. Large parts of the peripheral network with genuine traffic are
    left entirely unmonitored, which explains the high reconstruction error on those edges.

  \item \textbf{Betweenness:} Sensors follow the major routing corridors, forming a backbone
    spine. While core trunk links are covered, many end-point links and secondary paths fall
    outside the measured set.

  \item \textbf{Core-number:} Placement behaviour closely mirrors degree-based selection, as
    high core-number nodes tend to coincide with high-degree hubs in ISP topologies with limited
    k-core variation.

  \item \textbf{Random:} Sensor edges are scattered without structure. While spatial coverage
    is more uniform than hub-centric methods, the absence of any topological guidance means
    that the selected edges do not collectively span a meaningful algebraic subspace, leading
    to poor interpolation in the reconstruction step.
\end{enumerate}

\subsection{Performance Across Budget Levels}
\label{subsec:budget_sweep}
\begin{figure}[h]
  \centering
  \begin{tikzpicture}
  \begin{axis}[
      width=\columnwidth,
      height=6cm,
      xlabel={Sensor Budget (\% of edges monitored)},
      ylabel={Normalised Reconstruction Error},
      xtick={10,15,20,30,40},
      xticklabels={10\%,15\%,20\%,30\%,40\%},
      ymin=0.50, ymax=0.85,
      ytick={0.55,0.60,0.65,0.70,0.75,0.80,0.85},
      yticklabel style={/pgf/number format/fixed, /pgf/number format/precision=2},
      ymajorgrids=true,
      grid style={dashed, gray!30},
      axis line style={gray!60},
      tick style={gray!60},
      legend style={
          at={(0.5,1.05)},
          anchor=south,
          font=\tiny,
          draw=black,
          fill=white,
          legend columns=5,
      },
  ]

  \addplot[
      color={rgb,255:red,220;green,60;blue,60},
      mark=*, mark size=1.5pt, line width=1pt,
  ] coordinates {
      (10, 0.79) (15, 0.77) (20, 0.68) (30, 0.61) (40, 0.55)
  };
  \addlegendentry{ST-NDT}

  \addplot[
      color={rgb,255:red,60;green,80;blue,200},
      mark=square*, mark size=1.5pt, line width=1pt,
  ] coordinates {
      (10, 0.81) (15, 0.80) (20, 0.77) (30, 0.66) (40, 0.60)
  };
  \addlegendentry{Degree-based}

  \addplot[
      color={rgb,255:red,30;green,140;blue,100},
      mark=triangle*, mark size=1.5pt, line width=1pt,
  ] coordinates {
      (10, 0.81) (15, 0.80) (20, 0.75) (30, 0.80) (40, 0.78)
  };
  \addlegendentry{Betweenness}

  \addplot[
      color={rgb,255:red,140;green,80;blue,200},
      mark=diamond*, mark size=1.5pt, line width=1pt,
  ] coordinates {
      (10, 0.80) (15, 0.78) (20, 0.77) (30, 0.65) (40, 0.62)
  };
  \addlegendentry{Core-Number}

  \addplot[
      color={rgb,255:red,230;green,160;blue,30},
      mark=pentagon*, mark size=1.5pt, line width=1pt,
  ] coordinates {
      (10, 0.79) (15, 0.73) (20, 0.75) (30, 0.74) (40, 0.72)
  };
  \addlegendentry{Random}

  \end{axis}
  \end{tikzpicture}
  \caption{NRE vs.\ monitoring budget for all five methods. ST-NDT wins at 20\%--40\%; the performance gap widens with budget.}
\label{fig:error_vs_budget}
\end{figure}
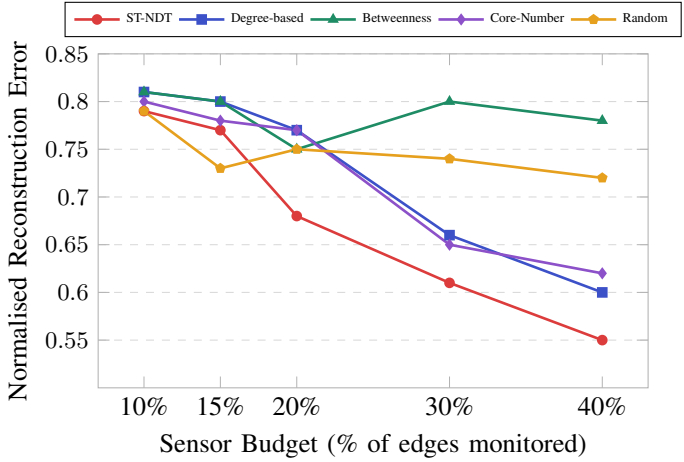
Figure~\ref{fig:error_vs_budget} depicts normalized reconstruction error (NRE) across monitoring budgets ranging from 10\% to 40\%. At low budgets (10\%--15\%), random placement achieves competitive or slightly lower NRE than the topological method, consistent with theoretical results showing that uniform random sampling can achieve near-optimal restricted isometry properties for certain signal classes. However, from 20\% onward, ST-NDT (Topological) is strictly superior at every budget level, achieving NRE = 0.684, 0.610, and 0.544 at 20\%, 30\%, and 40\% respectively. At 40\% coverage, this represents a $9.7\%$ improvement over the degree-based baseline (0.602) and a $19.9\%$ improvement over betweenness (0.743). Betweenness centrality shows the weakest scaling behaviour, with NRE falling by only $7.8$ percentage points across the full budget range compared to $26$ percentage points for the topological method, suggesting that bottleneck edges form a small, saturating set that provides diminishing marginal coverage gains. Averaged across all five budget levels, the topological method achieves the lowest mean NRE of 0.684, outperforming degree-based (0.729), core-number (0.731), random (0.687), and betweenness (0.780) baselines, confirming that the topological advantage grows consistently with monitoring resources.

\begin{table}[t]
\centering
\caption{Compatibility with emerging technologies and evolving standardization frameworks.}
\label{tab:3gpp_alignment}
\small
\setlength{\tabcolsep}{4pt}
\renewcommand{\arraystretch}{1.3}
\begin{tabular}{p{6.5cm} c}
\toprule
\textbf{6G / 3GPP Properties} & \textbf{ST-NDT (Topological)} \\
\midrule

\textbf{Network Digital Twin} \newline
3GPP TR 28.918 (Rel-18): Real-time digital replica of a live network for state estimation, monitoring and fault management.
\cite{Wu, mirzaei2023network}
& \checkmark \\

\textbf{Dense Mesh Topology} \newline
6G: Ultra-dense networks with cyclic redundancy paths require explicit handling of circular traffic flows.
\cite{saad2020vision}
& \checkmark \\

\textbf{Sparse AI/ML Monitoring} \newline
3GPP TR 38.843 (Rel-19): Minimum-overhead data collection with high-fidelity state reconstruction for RAN intelligence.
\cite{manohar2018data, 3gpp38843}
& \checkmark \\

\textbf{Semantic Communication (6G)} \newline
6G paradigm where only task-relevant information is transmitted, significantly reducing communication overhead and improving bandwidth efficiency \cite{10479470, 11395983} in large-scale network monitoring and digital twin synchronization.
& \checkmark \\

\bottomrule
\end{tabular}
\end{table}

Beyond outperforming baseline frameworks, the proposed ST-NDT framework aligns with the goals of 6G and 3GPP road-map, as summarized in Table~\ref{tab:3gpp_alignment}. Its mechanism directly supports the real-time state estimation requirements of 3GPP TR 28.918 by enabling continuous synchronization of live network states for monitoring in complex deployments. In addition, ST-NDT addresses the cyclic and redundant traffic patterns characteristic of ultra-dense 6G mesh topologies, where explicit handling of looped and cyclic flows is necessary to maintain stability and consistency. The framework also incorporates the principles of sparse and high-fidelity monitoring outlined in 3GPP TR 38.843 by prioritizing minimal-overhead data acquisition while preserving accurate state reconstruction for RAN intelligence. Finally, ST-NDT is consistent with the emerging 6G paradigm of semantic communication, in which only task-relevant information is transmitted, thereby reducing signaling overhead and improving spectral efficiency in large-scale network monitoring and digital twin synchronization. Collectively, these properties position ST-NDT as a unified and scalable framework for NGN management, bridging current 5G requirements with future 6G architectural principles.
\section{Conclusion}
\label{sec:conclusion}

In this paper, we presented a Sparse Topological Network Digital Twin (ST-NDT) framework that leverages Hodge theory on cell complexes to drive both sensor placement and signal reconstruction for sparse network monitoring. By representing the network as a cell complex and decomposing edge signals via the Hodge Laplacian, the framework achieved strong reconstruction performance across budgets of 10\%–40\% on a real-word ISP backbone topology, outperforming the benchmark frameworks. Future work will address temporal traffic dynamics via online re-optimization of the sensor set, and scalable solvers for deployment on large-scale topologies that characterize future 6G networks.

\section*{Acknowledgment}
This work is funded by the European Union. Views and opinions expressed are however those of the author(s) only and do not necessarily reflect those of the European Union or the Research Executive Agency. Neither the European Union nor the granting authority can be held responsible for them. Grant Agreement no. 101236523 AeroNet.
The authors would like to thank Prof. Sergio Barbarossa from University of Rome 1 Sapienza for the useful suggestions and scientific inspiration of this paper.

\bibliographystyle{unsrt}
\bibliography{references}

\end{document}